\documentclass{article}

\usepackage[preprint,nonatbib]{neurips_2020}

\usepackage[utf8]{inputenc} % allow utf-8 input
\usepackage[T1]{fontenc}    % use 8-bit T1 fonts
\usepackage{hyperref}       % hyperlinks
\usepackage{url}            % simple URL typesetting
\usepackage{booktabs}       % professional-quality tables
\usepackage{amsfonts}       % blackboard math symbols
\usepackage{nicefrac}       % compact symbols for 1/2, etc.
\usepackage{microtype}      % microtypography
\usepackage{comment}
\usepackage{graphicx}
\usepackage{multirow}
\usepackage[table]{xcolor}
\usepackage{collcell}
\usepackage{hhline}
\usepackage{pgf}

\title{A deep learning classifier for local ancestry inference}

\author{Matthew Aguirre \\
  Department of Biomedical Data Science \\
  Stanford University \\
  \texttt{magu@stanford.edu} \\
  
  \And  
  Jan Sokol \\
  Department of Biomedical Data Science \\
  Stanford University \\
  \texttt{jsokol@stanford.edu} \\

  \And
  Guhan Venkataraman \\
  Department of Biomedical Data Science \\
  Stanford University \\
  \texttt{guhan@stanford.edu} \\

  \And
  Alexander Ioannidis \\
  Department of Biomedical Data Science \\
  Stanford University \\
  \texttt{ioannid@stanford.edu}
}

\begin{document}

\maketitle

\begin{abstract}
   Local ancestry inference (LAI) identifies the ancestry of each segment of an individual's genome and is an important step in medical and population genetic studies of diverse cohorts. Several techniques have been used for LAI, including Hidden Markov Models and Random Forests. Here, we formulate the LAI task as an image segmentation problem and develop a new LAI tool using a deep convolutional neural network with an encoder-decoder architecture. We train our model using complete genome sequences from 982 unadmixed individuals from each of five continental ancestry groups, and we evaluate it using simulated admixed data derived from an additional 279 individuals selected from the same populations. We show that our model is able to learn admixture as a zero-shot task, yielding ancestry assignments that are nearly as accurate as those from the existing gold standard tool, RFMix.
\end{abstract}

\section{Introduction}

Ancestry inference refers to the task of assigning genetic ancestry labels given an individual's genomic sequence. Two forms are of note: global ancestry inference, which performs this task at an individual level, and local ancestry inference (LAI), which further segments an individual's genetic ancestry into assignments that can differ across the genome. 

Many models of ancestry inference have been proposed, spurring significant statistical methods development. The first model of global ancestry inference, STRUCTURE, was an independent discovery of Latent Dirichlet Allocation (LDA), and is still the preferred method for this task \cite{pritchard2000inference, blei2003latent, alexander2011enhancements}. Early models for local ancestry inference, such as HAPMIX, followed in the tradition of the Kingman coalescent and used extensions of a haplotype-based model of linkage disequilibrium due to Li and Stephens \cite{price2009sensitive, kingman1982coalescent, hudson1983properties, li2003modeling}. Other approaches to LAI have used Hidden Markov Models \cite{patterson2004methods} or discriminative models over windows of chromosomal sequence \cite{sankararaman2008estimating, maples2013rfmix}. These tools have been reviewed in greater depth by Liu, et. al. \cite{liu2013softwares}.

\begin{figure}[ht]
    \centering
    \includegraphics[width=\linewidth]{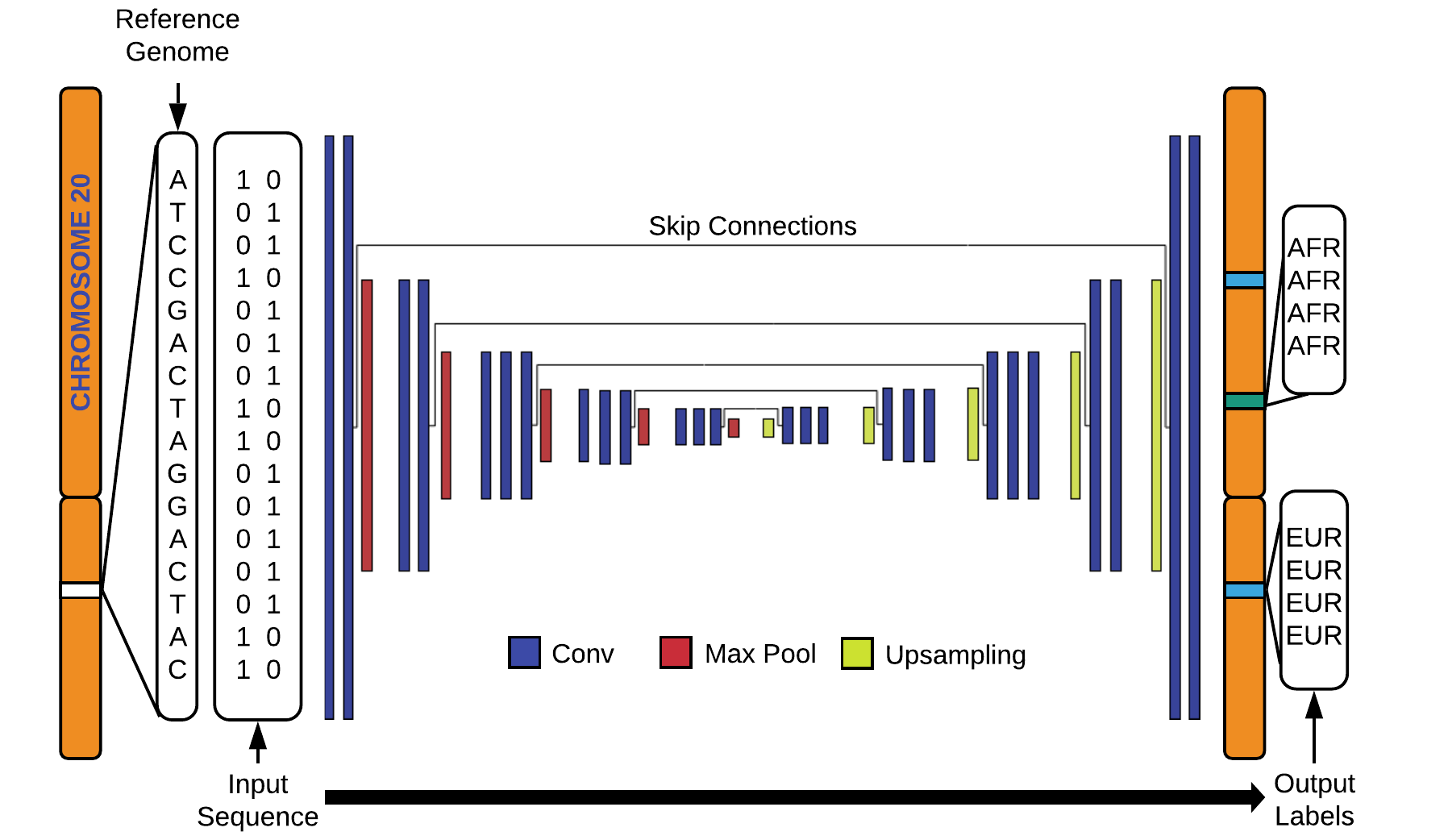}
    \caption{Model overview. Input is a two-channel bit vector that encodes whether a sample's maternal or paternal allele varies from the reference genome at a given site. Data are fed into a CNN based on the Segnet architecture \cite{badrinarayanan2017segnet}, which consists of several conv/maxpool and conv/upsampling blocks. The output is a segmentation of the input sequence, in which each letter is assigned to a genetic ancestry group.}
    \label{fig:overview}
\end{figure}

For most modern studies, however, LAI is a pure prediction problem in that model parameters are not of interest: the goal is the ancestry assignments themselves. Further, the LAI task can also be framed as image segmentation. In this formulation, the genome sequence is a one-dimensional image. Each variant site constitutes a pixel, with two complementary channels for reference and alternate alleles relative to a genome standard. Here, we describe a deep learning model for LAI using a convolutional neural network (CNN) designed for image segmentation, and benchmark it against the existing gold standard LAI tool, RFMix \cite{maples2013rfmix}. A schematic of our model is shown in \textbf{Figure \ref{fig:overview}}.

\section{Dataset}

We assembled a dataset of whole genome sequences from real individuals in the 1000 Genomes Project (1kG, $n=2,540$), Human Genome Diversity Project (HGDP, $n=929$), and Simons Genome Diversity Project (SGDP, $n=279$) \cite{10002015global, bergstrom2020insights, mallick2016simons}. These data were harmonized using standard bioinformatic tools such as PLINK and bcftools (to filter and merge genetic data), liftOver (to map variants to a common reference genome), and Beagle (to infer haplotypes from unphased samples) \cite{chang2015second, samtools, liftover, beagle}. 

We used ADMIXTURE, a tool which implements the STRUCTURE algorithm, to produce unsupervised genetic ancestry clusters, referencing metadata to interpret the resulting cluster assignments \cite{alexander2011enhancements}. At $K = 8$ clusters, we find genetic clusters consistent with African (AFR), East Asian (EAS), European (EUR), Native American (NAT), Oceanian (OCE), South Asian (SAS), San-Mbuti (SMB), and West Asian (WAS) ancestries. We removed individuals in the three least common cluster groups, as well as admixed individual (having $<99\%$ assignment to a single cluster), yielding a final dataset ($n=2,522$ haplotypes) of five continental ancestries (AFR, EAS, EUR, NAT, SAS), which we split into train, development (dev), and test sets (\textbf{Table~\ref{table:n}}). Related individuals were removed to avoid data leakage. For computational tractability, we subset the genetic data to include only biallelic (one alternate allele) single nucleotide polymorphisms (SNPs) on chromosome 20 that are present in the train set ($p=475,328$ variant positions).

\begin{table}[h!]
  \centering
  \begin{tabular}{l|lllll}
    \toprule
    Population & AFR & EAS & EUR & NAT & SAS \\
    \midrule
    Train & 289 & 389 & 118 & 51 & 135  \\
    Dev & 52 & 70 & 24 & 16 & 24  \\
    Test & 26 & 35 & 12 & 8 & 12  \\
    \bottomrule
  \end{tabular}
  \smallskip
  \caption{Number of individuals of each ancestral composition in the train, dev, and test sets.}
  \label{table:n}
\end{table}

% \vspace{-0.75 em}
To evaluate our models under realistic scenarios of admixture, we simulated the progeny of the real dev and test set individuals using RFMix \cite{maples2013rfmix}. We varied the number of generations of mixture and ancestral diversity of the founding members of the population so as to create a comprehensive set of benchmark datasets. For model selection (see Methods), we used a dev set consisting of 200 genotypes simulated by 10 generations of admixture across all five ancestries of dev set individuals.

\section{Methods}

SegNet is a convolutional encoder-decoder neural network architecture designed for semantic segmentation \cite{badrinarayanan2017segnet}. In its original instance SegNet consists of five ``down'' blocks containing two or three convolutional operators and a pooling layer, then five ``up'' blocks containing an up-sampling layer followed by two or three convolutional layers. Three SegNet hyperparameters are of interest: the number of blocks, the number of filters in the input layer, and the filter width. The number of filters increases by a factor of two in each down/up block, and filter width is fixed throughout. 

We conducted a grid search to select an optimal network architecture using these three hyperparameters. Filter width and filter count values were roughly spaced exponentially in the range $[2^1, 2^6]$, and the number of blocks was 3, 4, or 5. All models were trained using categorical cross-entropy loss, evaluated over genetic variants and individuals. The best performing model (with minimal dev set loss) had 4 blocks, with 16 filters with width 16. Complete results are available as a table on the project GitHub (\url{www.github.com/maguirre1/deepLAI/}). 

As baseline, we evaluated RFMix on the same test data. This method is fully described by Maples et. al. \cite{maples2013rfmix}; in brief, it uses several independent random forest classifiers to predict ancestry within chromosomal windows. Classifier output is then smoothed by a conditional random field (CRF) layer to yield final predictions. We used a version of RFMix (v2.03) from bioconda \cite{gruning2018bioconda}.

\section{Results}

\begin{figure}[ht]
    \centering
    \includegraphics[width=\linewidth]{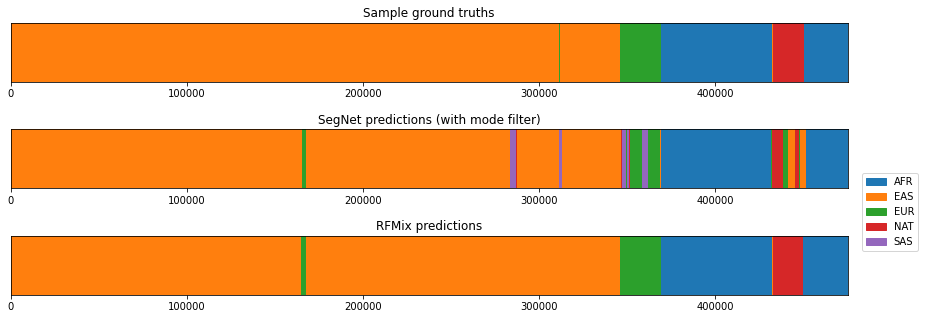}
    \caption{LAI Example. Genetic ancestry ground truth labels (top), with SegNet (middle) and RFMix predictions (bottom). $x$-axis values are position indexes. SegNet output is subject to a mode filter.}
    \label{fig:examples}
\end{figure}

In all, we trained about 50 models to select optimal hyperparameters. Several perform moderately well, achieving base-pair level accuracy above 80\% on the dev set (see Dataset), with the best model (see Methods) reaching 85.6\% accuracy. We also found that our model predictions are, in places, noisy (\textbf{Figure \ref{fig:examples}}). We therefore post-processed model output with a mode filter, replacing the output for each position with the most common prediction within $k$ positions on either side. We let $k=2000$, for an overall window size of 4000, and found that this increased the accuracy of our predictions by a few percentage points.

In an independent test set of individuals simulated under the same conditions as the dev set, our model achieves 88.2\% accuracy. Relative to the baseline method, RFMix, however, our model compares rather unfavorably (\textbf{Table \ref{table:test-set}}). The test set accuracy of RFMix, fit using the same training set as our model, is 97.2\%. While both models do quite well identifying regions of African (AFR) and East Asian (EAS) ancestries, the SegNet model fares much worse with the other populations. This is not surprising, as AFR and EAS are over-represented classes in the training set (\textbf{Table \ref{table:n}}). Moreover, the errors made by our model (e.g. confusion between EUR/SAS, and between NAT/EAS) tend to reflect the more recent common ancestry between those groups.

% \newpage % beware this hack!
\begin{table}[ht]
    \centering
    %    \parbox{.45\linewidth}{
    %    \centering
        \begin{tabular}{l|lllll}
            \toprule
            SegNet &    AFR &    EAS &    EUR &    NAT &    SAS \\
            \midrule
            AFR &  0.973 &  0.009 &  0.010 &  0.000 &  0.007 \\
            EAS &  0.005 &  0.949 &  0.006 &  0.002 &  0.039 \\
            EUR &  0.042 &  0.037 &  0.710 &  0.006 &  0.204 \\
            NAT &  0.004 &  0.182 &  0.010 &  0.739 &  0.066 \\
            SAS &  0.028 &  0.185 &  0.153 &  0.009 &  0.626 \\
            \bottomrule
        \end{tabular}
    %} \quad 
    %\parbox{.45\linewidth}{
    %    \centering
        \begin{tabular}{l|lllll}
            \toprule
            RFMix & AFR & EAS & EUR & NAT & SAS \\
            \midrule
            AFR & 0.981 & 0.004 & 0.013 & 0.000 & 0.002 \\ %\hhline{~*6{|-}|}
            EAS & 0.003 & 0.989 & 0.002 & 0.002 & 0.004 \\ %\hhline{~*6{|-}|} 
            EUR & 0.006 & 0.026 & 0.939 & 0.005 & 0.025 \\ %\hhline{~*6{|-}|}
            NAT & 0.001 & 0.011 & 0.029 & 0.954 & 0.004 \\ %\hhline{~*6{|-}|}
            SAS & 0.003 & 0.037 & 0.021 & 0.003 & 0.937 \\ %\hhline{~*6{|-}|}
            \bottomrule
        \end{tabular}
    %}
  \smallskip
  \caption{Model performance. Ground truth genotypes (rows) were generated by 10 generations of admixture between test set individuals. Confusion with output labels (columns) is normalized across rows. Diagonal entries are ancestry-specific sensitivity. All values rounded to thousandths.}
  \label{table:test-set}
\end{table}

% \vspace{-0.75 em}
Our model also generalizes across varied cohort demography. In another series of experiments, we generated 100 simulated individuals from 10 generations of admixture between founding populations of two ancestry groups at an unequally sampled (80:20) ratio. In Table \ref{table:test-80:20}, we highlight results from a cohort of 80\% African and 20\% South Asian ancestry. While RFMix achieves 98.4\% accuracy after fitting a new model, our SegNet model reaches 89.8\% without the need for retraining. We conducted similar experiments across all pairs of ancestry groups and find that model performance (i.e. class confusions from \textbf{Table \ref{table:test-set}}) is broadly consistent across the cohorts (full results on GitHub).

\begin{table}[ht]
    \centering
    %    \parbox{.47\linewidth}{
    %    \centering
        \begin{tabular}{l|lllll}
            \toprule
            SegNet &    AFR &    EAS &    EUR &    NAT &    SAS \\
            \midrule
            AFR &  0.989 &  0.001 &  0.004 &  0.000 &  0.006 \\
            SAS &  0.018 &  0.143 &  0.202 &  0.003 &  0.634 \\
            \bottomrule
        \end{tabular}
    %} \quad 
    %\parbox{.47\linewidth}{
    %    \centering
        \begin{tabular}{l|lllll}
            \toprule
            RFMix & AFR & EAS & EUR & NAT & SAS \\
            \midrule
            AFR & 0.989 & 0.001 & 0.009 & 0.000 & 0.001 \\ %\hhline{~*6{|-}|}
            SAS & 0.007 & 0.015 & 0.002 & 0.015 & 0.961 \\ %\hhline{~*6{|-}|}
            \bottomrule
        \end{tabular}
    %}
  \smallskip
  \caption{Model performance in 80/20 ancestry composition cohort. Ground truth genotypes (rows) are assigned output labels (columns), with values normalized across rows. Diagonal entries are ancestry-specific sensitivity. All values rounded to thousandths.}
  \label{table:test-80:20}
\end{table}
\vspace{-0.75 em}

\section{Discussion}

Here, we formulate the local ancestry inference task as a one-dimensional image segmentation problem, and develop a new LAI tool based on a deep convolutional neural network (SegNet). While there is a clear performance gap between the method we present here and the current gold standard, we foresee several ways to improve our model. One is data augmentation. This includes incorporating admixed genotypes during training, which may be generated on the fly within each epoch during training (as RFMix does), or which may be pre-computed. Another is adding data ``channels'' for biologically relevant quantities -- such as recombination rate between sites (also considered by RFMix), or population-specific allele frequencies for each input variant. Architectural changes, including adding a conditional random field output layer to smooth output predictions, could also be helpful. Finally, more extensive use of regularization techniques like dropout may also improve performance, as most genetic variants are not informative with respect to ancestry. 

Though work remains to make our model viable, we note that our approach is not without benefits or novelty. First, while the models in this work are not very resource-hungry (all run to completion in under two days on a single NVIDIA RTX 2080 Ti), the ``train once, use anywhere'' paradigm is immensely important for analyses of large or growing datasets \cite{noto2020ancestry}. The portability of the model allows ancestry learned on a privacy-protected or proprietary dataset to be applied by external users lacking access, something that is not possible with RFMix. Second, it warrants mention that deep learning techniques are underutilized in population genetics, and many popular genomics tools are not optimized for graphical processing resources. We believe there is significant room for methods development in this space, particularly for purely predictive tasks such as LAI. We anticipate this will be of interest as the size and scope of genomic data continue to grow.

\section{Broader Impacts}

Several commercial entities offer generic ancestry inference as a direct-to-consumer enterprise, which has remained largely uncontroversial, aside from data privacy concerns and occasional unexpected family discoveries. However, we must acknowledge that the field of genetics -- in particular, genetic studies of differences between individuals of diverse backgrounds -- has had a problematic past with respect to data collection (e.g. participant consent) and to propagating systems of racial hierarchy and racism (e.g. eugenics). Even today, there remains significant risk that certain individuals will promote misreadings of genetic studies to further personal or political agendas. In this respect, let us be very clear: we believe that race, as a social construct, has no biological basis and therefore no meaningful place in population genetic studies like this one.

Population genetic studies, using tools such as the ones we propose, can reveal a great deal about our origins and history as a species: our migrations, our interactions, and our families. For some, these stories may be a source of pride, or a tool to reclaim a lost, colonized past, while for others, they may be reminders of times where our ancestors have been less kind. These realities do not negate the ways we choose to describe ourselves in the present, and they do not foretell the type of people we may become.

\small

\bibliographystyle{unsrt}
\bibliography{references}

\end{document}